\documentclass[12pt]{amsart}
\usepackage{amssymb,latexsym,amsmath,amsthm,enumitem,mathrsfs,geometry}
\usepackage{fullpage}
\usepackage{fancyhdr}
\usepackage{xcolor}
\usepackage{bbm}
\usepackage{stmaryrd}
\usepackage{mathrsfs}
\usepackage{hyperref}
\usepackage{lineno}
\usepackage{diagbox}
\usepackage{array}
\usepackage{tikz}
\usetikzlibrary{arrows}
\usetikzlibrary{positioning}
\usepackage{float}
\usepackage{comment}
\usepackage{mathtools}
\usepackage{cleveref}

\usepackage{multirow}

\newtheorem{theorem}{Theorem}[section]
\newtheorem*{theorem*}{Theorem}

\newtheorem*{remark*}{Remark}

\begin{document}

\numberwithin{equation}{section}

\title{On invariant solutions of linear time-fractional diffusion-wave equations with variable coefficients}

\author[S. Adiya]{Sodbaatar Adiya}
\address{Department of Mathematics, National University of Mongolia, Ulaanbaatar, Mongolia}  
\email{sodoomath@gmail.com}  

\author[Kh. Dorjgotov]{Khongorzul Dorjgotov} 
  \address{Department of Mathematics, National University of Mongolia, Ulaanbaatar, Mongolia}  
  \email{pilpalpil@gmail.com} 

  \author[B. Gombodorj]{Bayarmagnai Gombodorj} 
  \address{Department of Mathematics, National University of Mongolia, Ulaanbaatar, Mongolia}  
  \email{bayarmagnai@smcs.num.edu.mn} 

    \author[H. ochiai]{Hiroyuki Ochiai} 
  \address{Institute of Mathematics for Industry, Kyushu University, Fukuoka, Japan}  
  \email{ochiai@imi.kyushu-u.ac.jp}

    \author[U. Zunderiya]{Uuganbayar Zunderiya$^*$} 
  \address{Department of Mathematics, National University of Mongolia, Ulaanbaatar, Mongolia}  
  \email{zunderiya@gmail.com}

\thanks{$^*$Corresponding Author: U. Zunderiya}


\begin{abstract} We study invariant solutions of a certain class of time-fractional diffusion-wave equations with variable coefficients via Lie symmetry analysis. In physics, the fractional diffusion equation describes transport dynamics that are governed by anomalous diffusion while the fractional wave equation describes oscillations and wave propagation in various physical systems. In order to obtain exact invariant solutions of these equations, we firstly determine infinitesimal symmetries with respect to the variable coefficients of the equations. With the help of these symmetries, we then find solutions in terms of Mittag-Leffler functions, generalized Wright functions and Fox H-functions.
\end{abstract}

\maketitle

\section{Introduction}\label{sec:introduction}
Fractional diffusion-wave equations (FDWEs) have accurate descriptions of many physical processes including the modeling of anomalous diffusive systems, wave propagation phenomena and problems in electromagnetic theory.
For example, many electromagnetic and acoustic problems in physics can be modeled accurately and efficiently using FDWE \cite{nigmatullin2,nigmatullin1}. 
The fractional diffusion equation was introduced in physics by studying diffusion processes in media with fractal geometry \cite{nigmatullin1}, while the fractional wave equation was introduced by investigating the propagation of mechanical diffusive waves in visco-elastic media which exhibit a power-law creep \cite{Mainardi4}. Further, time FDWEs were applied to the propagation of stress waves in visco-elastic media relevant to acoustics and seismology \cite{Mainardi5}. Recently, FDWEs were used in order to obtain the generalized formula for retarded potentials in electrodynamics with Maxwell equations containing fractional time derivatives \cite{pskhu1}. 
For more research related to fractional differential equations and their applications in various fields of physics and other sciences, we recall the works \cite{glockle,kilbas1,Kiryakova,Metz,oldh,Pod,sakamoto1,sun1,tarasov1,tarasov2,baleanu1,baleanu2} and references therein.

In this study, we consider a class of linear time FDWEs with variable coefficients of the following form
\begin{equation}\label{1}
\partial^{\alpha}_t u(x,t)=a(x)^2u(x,t)_{xx}+b(x)u(x,t)_x, \quad a(x)\neq 0,
\end{equation}
where time derivative order $\alpha$ is a positive real number, $a(x)$ and $b(x)$ are sufficiently differentiable functions.
The Riemann-Liouville fractional differentiation is defined as
\begin{align}\label{defrld}
\partial^{\alpha}_t u(x,t):=\left\{\begin{array}{ll}
\frac{\partial^n u}{\partial t^n},&\mbox{ for }\alpha=n\in\mathbb{N},\\
\frac{1}{\Gamma(n-\alpha)}\frac{\partial^n}{\partial t^n}  \int_{0}^{t}\frac{u(x,s)}{(t-s)^{\alpha-n+1}}ds,&\mbox{ for } \alpha\in(n-1,n),\text{ } n\in\mathbb{N}.
\end{array}
\right.
\end{align}
By employing Lie symmetry analysis, we investigate symmetries and obtain invariant solutions to Eq.~(\ref{1}). Our results contain previously known solutions as their particular cases in a sense of time derivative order $\alpha$ and the coefficients $a(x)$ and $b(x)$. For example, symmetries and invariant solutions to Eq.~(\ref{1}) when $\alpha=2$ and $b(x)=0$, are studied in \cite{bluman1} as well as in \cite{lie1,ovs,gungor1,gungor2} when $\alpha=1$.
For $\alpha > 1$, symmetries and invariant solutions to Eq.~(\ref{1}) with constant coefficients i.e. $a(x)=\text{const}$ and $b(x)=0$, are obtained in \cite{Luchko,Luchko1}. 
For $0<\alpha<1$, invariant solutions of Eq.~(\ref{1}) with $a(x)=ax^m$, $a>0$ and $b(x)=bx^{m-1}$ are given in \cite{Metzler1} by using Laplace transformations.

In the following sections of this paper, we first obtain the Lie point symmetries of Eq.~(\ref{1}) depending on the coefficient functions. We then explicitly derive invariant solutions corresponding to infinitesimal symmetries.

\section{Group classification of time FDWEs}
\label{intro}

We look for invariant solutions of time FDWEs described by Eq.~(\ref{1}) under Lie point symmetry. The infinitesimal and extended infinitesimal generators of Eq.~(\ref{1}) are given as
\begin{equation*}
X=\xi(x,t,u)\partial_x+\tau(x,t,u)\partial_t+\eta(x,t,u)\partial_u 
 \quad \text{and} 
\end{equation*}
\begin{equation}
\label{eq:X}   
X^{*}=X + \mu^{(1)}\partial_{u_x}+\mu^{(2)} \partial_{ u_{xx}}+\mu^{(\alpha)} \partial_{\partial^\alpha_t u},
\end{equation}
respectively in \cite{Gaz3},
where 
\begin{eqnarray*}
\mu^{(1)}&=&\eta_x+(\eta_u-\xi_x)u_x-\tau_xu_t-\xi_u u_x^2-\tau_uu_xu_t,\\
\mu^{(2)}&=&\eta_{xx}+(2\eta_{xu}-\xi_{xx})u_x-\tau_{xx}u_t+(\eta_{uu}-2\xi_{xu})u_x^2-2\tau_{xu}u_xu_t
-\xi_{uu}u_x^3\\
& & -\tau_{uu}u_x^2u_t+(\eta_u-2\xi_x)u_{xx} -2\tau_xu_{xt}-3\xi_uu_{xx}u_x
-\tau_u u_{xx}u_t-2\tau_u u_{xt} u_x,\\
\mu^{(\alpha)}&=&\partial^\alpha_t \eta+(\eta_u-\alpha D_t\tau)\partial^\alpha_t u - u\partial^\alpha_t \eta_u\\
&& +\sum_{n=1}^\infty \left[\binom{\alpha}{n} \partial^n_t \eta_u- \binom{\alpha}{n+1}D_t^{n+1}\tau \right] \partial^{\alpha-n}u- \sum_{n=1}^\infty \binom{\alpha}{n}D_t^n\xi \partial^{\alpha -n}u_x+\mu,
\end{eqnarray*}
with
\begin{equation*}
\mu=\sum_{n=2}^\infty\sum_{m=2}^n\sum_{k=2}^m\sum_{r=0}^{k-1}\binom{\alpha}{n}\binom{n}{m}\binom{k}{r}\frac{1}{k!}\frac{t^{n-\alpha}}{\Gamma(n+1-\alpha)}\left(-u\right)^r\frac{\partial^m}{\partial t^m}\left(u^{k-r}\right)\frac{\partial^{n-m+k}\eta}{\partial t^{n-m}\partial u^k},
\end{equation*}
\begin{equation*}
\binom{\alpha}{n}=\frac{(-1)^{n-1}\alpha\Gamma(n-\alpha)}{\Gamma(1-\alpha)\Gamma(n+1)}
\quad \text{and} \quad 
D_t = \frac{\partial}{\partial t}+u_t\frac{\partial}{\partial u}+u_{xt}\frac{\partial}{\partial u_x}+\cdots.
\end{equation*}
It should be noted that $\mu=0$ when $\mu$ is linear in $u$.
Moreover, the initial condition becomes  
\begin{equation}\label{7}
\tau(x,t,u)|_{t=0}=0
\end{equation}
due to the fixed lower limit in the integral in (\ref{defrld}).

The infinitesimal symmetries for group-invariant solutions of Eq.~(\ref{1}) should satisfy the determining equation
\begin{equation}
\label{eq:detX}
    \left.X^{*}\left(\frac{\partial^\alpha u}{\partial t^\alpha}-a(x)^2u_{xx}-b(x)u_x\right)\right|_{(\ref{1})}=0.
\end{equation}
By substituting (\ref{eq:X}) into (\ref{eq:detX}) and equating the coefficients of various linearly independent variables to zero, we obtain the following system of equations:
\begin{eqnarray}
\label{eq:d1}
&& \tau_x=0, \quad \tau_u=0, \quad \xi_u=0,\quad \binom{\alpha}{n}D_t^n\xi=0, ~\forall n\in  \mathbb{N},\\
\label{eq:dd1}
&&a(x)^2(2\xi_x-\alpha D_t\tau)-2a(x)a(x)'\xi=0,\\
\label{eq:d7}
&& -a(x)^2\eta_{uu}=0,\quad \binom{\alpha}{n} \partial^n_t \eta_u-
\binom{\alpha}{n+1} D_t^{n+1}\tau=0,~\forall n\in  \mathbb{N},\\
\label{eq:d2}
&&\partial^\alpha_t \eta-
u\partial^\alpha_t \eta_u-a^2\eta_{xx}-b(x)\eta_x+\mu=0,\\
\label{eq:d6}
&& b(x)\left(\xi_x-\alpha D_t\tau\right)-a^2(2\eta_{xu}-\xi_{xx})-b(x)'\xi=0.
\end{eqnarray}
The equations in (\ref{eq:d1}) and (\ref{eq:dd1}) along with the condition (\ref{7}) yield us
\begin{equation}
\label{eq:xi}
  \tau(t)=\frac{s_1}{\alpha}t\quad\mbox{and}\quad \xi(x)=a(x) \left( s_2+
    \frac{s_1}{2}\int_\beta^x \frac{dr}{a(r)} \right),
\end{equation}
where $s_1,s_2\in \mathbb{R}$ and $\beta$ is a suitable real number.
By solving the equations in (\ref{eq:d7}) and (\ref{eq:d6}), we have $\mu=0$ and
\begin{equation}
\label{eq:eta}
   \eta=\left[-\frac{1}{2}\left(\frac{b(x)}{a(x)}-a(x)'\right)\left( s_2+
    \frac{s_1}{2}\int_\beta^x \frac{dr}{a(r)} \right)+s_3\right]u+d(x,t),
\end{equation}
where $d(x,t)$ is a function depending on $x$ and $t$ only. Furthermore, if we substitute (\ref{eq:xi}) and (\ref{eq:eta}) into (\ref{eq:d2}), we see that $d(x,t)$ is an arbitrary solution of Eq.~(\ref{1}) and
\begin{multline}\label{last1}
2\left(c(x)'a(x)+\frac{1}{2}c(x)^2\right)'s_2\\
=-\left[\left(c(x)'a(x)+\frac{1}{2}c(x)^2\right)'\int_\beta^x \frac{dr}{a(r)}+2\left(c(x)'a(x)+\frac{1}{2}c(x)^2\right)\frac{1}{a}\right]s_1,
\end{multline}
where $c(x)=\frac{b(x)}{a(x)}-a(x)'$. Thus, for the arbitrary functions  $a(x)$ and $b(x)$, Eq.~(\ref{1}) has the following infinitesimal symmetries
$$X_{d(x,t)}=d(x,t)\partial_u \quad\mbox{ and }\quad X_1=u\partial_u,$$
where d(x,t) is an arbitrary solution of Eq.~(\ref{1}).

By analyzing the relation (\ref{last1}), all of the infinitesimal symmetries of Eq.~(\ref{1}) that satisfy the determining Eq.~(\ref{eq:detX}), are given by the following Table \ref{table:1}.

\newpage

\begin{table}[h!]
    \centering
   \caption{ $\begin{array}{c} b(x) = a(x)c(x) + a(x)a(x)',~ \omega(x)=\int_\beta^x \frac{dr}{a(r)} +\lambda_1,\\
    (\text{where } \lambda_1\in \mathbb{R} \text{ and } \beta \text{ is a suitable real number})
    \end{array}$}
    \label{table:1}
    \begin{tabular}{|c|c|l|}
\hline
Cases&$c(x)$ & \multicolumn{1}{c|}{$X$}\\
 \hline 
\multirow{2}{*}{1} &
 any sufficiently differentiable&$X_{d(x,t)}=d(x,t)\partial_u,$\\ 
&function except cases 2-8&$X_1=u\partial_u$\\
 \hline 
\multirow{3}{*}{2} &
 \multirow{3}{*}{$\dfrac{\left(\ln \omega(x)+\lambda_2\right)}{\omega(x)\left(\ln \omega(x)+\lambda_2-2\right)},$} & 
$ X_{d(x,t)}=d(x,t)\partial_u,$\\ 
&& $X_1=u\partial_u,$\\
&$\lambda_2\in \mathbb{R}$& $X_2=a(x)\omega(x)\partial_x+\frac{2}{\alpha}t\partial_t-\dfrac{c(x)}{2}u\partial_u$\\
 \hline
 \multirow{3}{*}{3} &
 \multirow{2}{*}{$\dfrac{\left(\lambda_2+1\right)\bigl(\lambda_3\omega(x)^{\lambda_2}-\lambda_2+1\bigl)}{\omega(x)\left(\lambda_3\omega(x)^{\lambda_2}+\lambda_2+1\right)},$} & 
$ X_{d(x,t)}=d(x,t)\partial_u,$\\ 
&& $X_1=u\partial_u,$\\
&$\lambda_2,\lambda_3\in \mathbb{R},~\lambda_2\neq -1$& $X_3=a(x)\omega(x)\partial_x+\frac{2}{\alpha}t\partial_t-\dfrac{c(x)}{2}u\partial_u$\\
 \hline
 \multirow{3}{*}{4} &
 \multirow{2}{*}{$\dfrac{\bigl(\lambda_2^2+1\bigr)}{\omega(x)\left(\lambda_2 \tan\left(\frac{\lambda_2}{2} \ln \omega(x)+\lambda_3\right)+1\right)}$,} & 
$ X_{d(x,t)}=d(x,t)\partial_u,$\\ 
&& $X_1=u\partial_u,$\\
&$\lambda_2,\lambda_3\in \mathbb{R},$ $\lambda_2\neq 0$ & $X_4=a(x)\omega(x)\partial_x+\frac{2}{\alpha}t\partial_t-\dfrac{c(x)}{2}u\partial_u$\\
 \hline
 \multirow{3}{*}{5} &
 \multirow{3}{*}{$\dfrac{\lambda_2\left(\epsilon e^{\lambda_2\omega(x)}-1\right)}{\epsilon e^{\lambda_2\omega(x)}+1},$} & 
$ X_{d(x,t)}=d(x,t)\partial_u,$\\ 
&& $X_1=u\partial_u,$\\
&$\lambda_2\in \mathbb{R},~\lambda_2\neq 0,~\epsilon=\pm1$& $X_5=a(x)\partial_x-\dfrac{c(x)}{2}u\partial_u$\\
 \hline
 \multirow{3}{*}{6} &
 \multirow{3}{*}{$\lambda_2\tan\left(-\frac{\lambda_2}{2}\omega(x)\right),$} & 
$ X_{d(x,t)}=d(x,t)\partial_u,$\\ 
&& $X_1=u\partial_u,$\\
&$\lambda_2\in \mathbb{R}, \lambda_2\neq 0$& $X_6=a(x)\partial_x-\dfrac{c(x)}{2}u\partial_u$\\
 \hline
  \multirow{4}{*}{7} &
 \multirow{4}{*}{$0$} & 
$ X_{d(x,t)}=d(x,t)\partial_u,$\\ 
&& $X_1=u\partial_u,$\\
&& $X_7=a(x)\omega(x)\partial_x+\frac{2}{\alpha}t\partial_t,$\\
&& $X_8=a(x)\partial_x$\\
 \hline
 \multirow{4}{*}{8} &
 \multirow{4}{*}{$\dfrac{2}{\omega(x)}$} & 
$ X_{d(x,t)}=d(x,t)\partial_u,$\\ 
&& $X_1=u\partial_u,$\\
&& $X_7=a(x)\omega(x)\partial_x+\frac{2}{\alpha}t\partial_t,$\\
&& $X_9 =a(x)\partial_x-\dfrac{c(x)}{2}u\partial_u$\\
 \hline
    \end{tabular}
\end{table}

In cases 2-8 in Table \ref{table:1}, Eq.~(\ref{1}) becomes 
\begin{equation}\label{new2}
\partial^{\alpha}_t u(x,t)=a(x)^2u(x,t)_{xx}+\left(a(x)\overline{c}(\omega(x)) + a(x)a(x)'\right)u(x,t)_x, \quad a(x)\neq 0,
\end{equation}
where $\overline{c} (\omega(x))=c(x)$. By the transformation $\omega=\int_\beta^x \frac{dr}{a(r)} +\lambda_1$, Eq.~(\ref{new2}) is then reduced the equation
\begin{equation}\label{eqnew1}
\partial^{\alpha}_t u(\omega,t)=u(\omega,t)_{\omega\omega}+\overline{c}(\omega)u(\omega,t)_\omega.
\end{equation}
Hence, we are now able to formulate the
following theorem.
\begin{theorem}
\label{Thm1}
All of the infinitesimal symmetries of Eq.~(\ref{eqnew1}) are given by the following cases:

\begin{enumerate}
\item for any sufficiently differentiable $\overline{c}(\omega)$,
$$X_{d(\omega,t)}=d(\omega,t)\partial_u \quad\mbox{ and }\quad X_1=u\partial_u,$$
where $d(\omega,t)$ is an arbitrary solution of Eq.~(\ref{eqnew1}); 

\item for $\overline{c}(\omega)=\dfrac{\left(\ln \omega+\lambda_2\right)}{\omega\left(\ln \omega+\lambda_2-2\right)}$, $\lambda_2\in \mathbb{R}$, 
\begin{eqnarray*}
& &X_{d(\omega,t)},\quad X_1\quad\mbox{ and }\\
& &X_2=\omega\partial_\omega+\frac{2}{\alpha}t\partial_t-\dfrac{\ln \omega+\lambda_2}{2\left(\ln \omega+\lambda_2-2\right)}u\partial_u;
\end{eqnarray*}
\item for $\overline{c}(\omega)=\dfrac{\left(\lambda_2+1\right)\bigl(\lambda_3\omega^{\lambda_2}-\lambda_2+1\bigl)}{\omega\left(\lambda_3\omega^{\lambda_2}+\lambda_2+1\right)}$, $\lambda_2,\lambda_3\in \mathbb{R}$, $\lambda_2\neq -1$,
\begin{eqnarray*}
&&X_{d(\omega,t)},\quad X_1\quad\mbox{ and }\\
&&X_3=\omega\partial_\omega+\frac{2}{\alpha}t\partial_t-\dfrac{\left(\lambda_2+1\right)\left(\lambda_3\omega^{\lambda_2}-\lambda_2+1\right)}{2\left(\lambda_3\omega^{\lambda_2}+\lambda_2+1\right)}u\partial_u;
\end{eqnarray*}
\item for $\overline{c}(\omega)=\dfrac{\bigl(\lambda_2^2+1\bigr)}{\omega\left(\lambda_2 \tan\left(\frac{\lambda_2}{2} \ln \omega+\lambda_3\right)+1\right)}$, $\lambda_2,\lambda_3\in \mathbb{R}$, $\lambda_2\neq 0$,
\begin{eqnarray*}
&&X_{d(\omega,t)},\quad X_1\quad\mbox{ and }\\
&&X_4=\omega\partial_\omega+\frac{2}{\alpha}t\partial_t-\dfrac{\bigl(\lambda_2^2+1\bigr)}{2\left(\lambda_2\tan\left(\frac{\lambda_2}{2} \ln \omega+\lambda_3\right)+1\right)}u\partial_u;
\end{eqnarray*}

\item for $\overline{c}(\omega)=\dfrac{\lambda_2\left(\epsilon e^{\lambda_2\omega}-1\right)}{\epsilon e^{\lambda_2\omega}+1}$, $\lambda_2\in \mathbb{R},~\lambda_2\neq 0,~\epsilon=\pm1,$
$$X_{d(\omega,t)},\quad X_1\quad\mbox{ and }\quad X_5=\partial_\omega-\dfrac{\lambda_2\left(\epsilon e^{\lambda_2\omega}-1\right)}{2\left(\epsilon e^{\lambda_2\omega}+1\right)}u\partial_u;$$

\item for $\overline{c}(\omega)=\lambda_2\tan\left(-\frac{\lambda_2}{2}\omega\right)$, $\lambda_2\in \mathbb{R}$, $\lambda_2\neq 0$,
$$X_{d(\omega,t)},\quad X_1\quad\mbox{ and }\quad X_6=\partial_\omega-\frac{\lambda_2}{2}\tan\left(-\frac{\lambda_2}{2}\omega\right)u\partial_u;$$

\item for $\overline{c}(\omega)=0$, 
$$X_{d(\omega,t)},\quad  X_1, \quad X_7=\omega\partial_\omega+\frac{2}{\alpha}t\partial_t \quad \mbox{ and } \quad
X_8=\partial_\omega;$$

\item for $\overline{c}(\omega)=\dfrac{2 }{\omega}$, 
$$X_{d(\omega,t)},\quad  X_1, \quad X_7\quad  \mbox{ and } \quad X_9 =\partial_\omega -\frac{1}{\omega} u \partial_u.$$

\end{enumerate}
\end{theorem}

Table \ref{table:1} and Theorem~\ref{Thm1} gives us a complete group classification of Eq.~(\ref{1}). In the next section, we express invariant solutions as solutions of so-called reduced fractional ordinary differential equations. Finally, we obtain exact invariant solutions to Eq.~(\ref{eqnew1}) by using similarity transformations.

\section{Invariant solutions of time FDWEs}
\label{invariantsol}
In this section, we look for invariant solutions that remain unaltered under all transformations in an one-parameter group $G_a$ generated by an infinitesimal symmetry of Eq.~(\ref{1}). From \cite{symmetrybook}, it is well known that the solution $u(x,t)$ of Eq.~(\ref{1}) is an invariant of the symmetry group $G_a$ with infinitesimal symmetry $X=\xi\partial_x+\tau\partial_t+\eta\partial_u$
if and only if it solves the invariant surface condition $\xi u_x+\tau u_t-\eta=0.$

We recall the following special functions and their properties in preparation for introducing exact invariant solutions of Eq.~(\ref{1}). The Fox H-function (e.g. in \cite{glockle,Kiryakova,Hnom}) is defined by means of the Mellin-Barnes type contour integral
\begin{equation}
\label{int1}
H_{p,q}^{m,l}\left[z\biggr\vert\begin{array}{c}
(a_i, \alpha_i)_{1,p}\\
(b_j, \beta_j)_{1,q}
\end{array}\right]=\frac{1}{2\pi i}\int_{L}\frac{\prod\limits_{j=1}^m\Gamma(b_j-\beta_js)\prod\limits_{i=1}^l\Gamma(1-a_i+\alpha_is)}{\prod\limits_{i=l+1}^p\Gamma(a_i-\alpha_is)\prod\limits_{j=m+1}^q\Gamma(1-b_j+\beta_js)}z^{s}d s,
\end{equation}
for $z\in\mathbb{C}\setminus\{0\},$ where $m,l,p,q\in\mathbb{N}_0=\{0,1,2,\ldots\}$, $(m,l)\neq (0,0),$ $\alpha_i, \beta_j\in\mathbb{R}_+$, $a_i,b_j\in\mathbb{R}$ ($i=1,\ldots, p;j=1,\ldots,q$). Here $L$ is a suitable contour from $\gamma-i\infty$ to $\gamma+i\infty$, where $\gamma$ is a real number. The integral in (\ref{int1}) converges if the following conditions are met 
\begin{equation*}
\rho=\sum_{i=1}^{l}\alpha_i-\sum_{i=l+1}^{p}\alpha_i+\sum_{j=1}^{m}\beta_j-\sum_{j=m+1}^{q}\beta_j>0\quad\mbox{and}\quad |\arg z|<\frac{\pi\rho}{2}. 
\end{equation*}
The Fox H-function vanishes for large $z$ because 
\begin{equation*}
\label{eq8}
H_{p,q}^{m,0}[z]\approx O\left(\exp\left(-\nu z^{\frac{1}{\nu}}\mu^{\frac{1}{\nu}}\right)z^{\frac{2\delta+1}{2\nu}}\right),
\end{equation*}
where $\mu=\prod\limits_{i=1}^p\alpha_i^{\alpha_i}\prod\limits_{j=1}^q\beta_j^{-\beta_j},$ $\delta=\sum\limits_{j=1}^q b_j-\sum\limits_{i=1}^p a_i+\frac{p-q}{2}$  and
$\nu=\sum\limits_{j=1}^{q}\beta_j-\sum\limits_{i=1}^{p}\alpha_i>0$.

The generalized Wright function (e.g. in \cite{Luchko1,kilb2005}) is defined as
\begin{eqnarray}
\label{ub3}
{}_p\Psi_q\left[z\left|\begin{array}{c}
(a_i,\alpha_i)_{1,p}\\
(b_j,\beta_j)_{1,q}
\end{array}\right.\right] & = & \sum_{k=0}^\infty\frac{\prod\limits_{i=1}^p\Gamma(a_i+\alpha_i k)}{\prod\limits_{j=1}^q\Gamma(b_j+\beta_j k)}\frac{z^k}{k!},
\end{eqnarray}
for $z\in\mathbb{C},$ $p,q\in\mathbb{N}_0,$ $a_i,b_j\in\mathbb{C}$ and $\alpha_i, \beta_j\in\mathbb{R}\setminus\{0\}$ ($i=1,\ldots,p; j=1,\ldots,q$). 

If $\Delta=\sum\limits_{j=1}^q \beta_j-\sum\limits_{i=1}^p \alpha_i>-1$ or $\Delta=-1,$ then the series in (\ref{ub3}) is absolutely convergent for $z\in \mathbb{C}$ or $|z|<\prod\limits_{i=1}^p|\alpha_i|^{-\alpha_i}\prod\limits_{j=1}^q|\beta_j|^{\beta_j}$, respectively. Moreover, the Mittag-Leffler, Wright and Gauss hypergeometric functions can be expressed in terms of the generalized Wright functions, respectively, as
\begin{equation*}
E_{\alpha,\beta}(z) = {}_1\Psi_1\left[z\left|\begin{array}{c}
(1,1)\\
(\beta,\alpha)
\end{array}\right.
\right],\quad 
\Psi\left(z;\alpha,\beta\right) ={}_0\Psi_{1}\left[z\left|\begin{array}{c}
-\\
(\beta,\alpha)
\end{array}\right.\right] \quad\mbox{and}
\end{equation*}
$${}_2F_1\left(\begin{array}{c}
A,B\\
C
\end{array};z\right)=\frac{\Gamma(C)}{\Gamma(A)\Gamma(B)}{}_2\Psi_{1}\left[z\left|\begin{array}{c}
(A,1),(B,1)\\
(C,1)
\end{array}\right.\right].$$

We now give  invariant solutions of Eq.~(\ref{1}) in terms of the above mentioned special functions. These invariant solutions correspond to some infinitesimal symmetries of the Lie algebras generated by the infinitesimal symmetries listed in Theorem~\ref{Thm1}. 

\subsection{The generator $V_{1}=\left(s+\frac{1}{2}\right)X_1+X_2$, $s\in \mathbb{R}$} \label{subsec:V11}
\textcolor{white}{.}\\ 
The characteristic equation for $V_{1}$ is
$$
\frac{d\omega}{\omega}=\frac{\alpha dt}{2t}=\frac{2du}{\left(2s+1-\dfrac{\ln \omega+\lambda_2}{\ln \omega+\lambda_2-2}\right)u},
$$
which yields the following similarity transformation and similarity variable
\begin{equation}
\label{eq:V11simtrans}
    u(\omega,t)=\frac{\omega^s\varphi(z)}{\ln\omega+\lambda_2-2} \quad \text{and}\quad z=\omega^{-\frac{2}{\alpha}}t.
\end{equation}
By substituting it into Eq.~(\ref{eqnew1}) with $\overline{c}(\omega)=\dfrac{\left(\ln \omega+\lambda_2\right)}{\omega\left(\ln \omega+\lambda_2-2\right)}$, we have 
\begin{equation}
\label{eq:V11red}
\frac{d^\alpha}{dz^\alpha}\varphi(z) = s^2\varphi(z)+\frac{4}{\alpha}\left(\frac{1}{\alpha}-s\right)z\varphi(z)'+\frac{4}{\alpha^2}z^2\varphi(z)''.
\end{equation}
We now apply Propositions 3.1 and 3.2 in \cite{ourarticle3} in order to give solutions to Eq.~(\ref{eq:V11red}) and then substitute these solutions into (\ref{eq:V11simtrans}). As a result, we obtain explicit invariant solutions of Eq.~(\ref{eqnew1})  with $\overline{c}(\omega)=\dfrac{\left(\ln \omega+\lambda_2\right)}{\omega\left(\ln \omega+\lambda_2-2\right)}$ as follows
\begin{equation}
\label{uuu1}
    u(\omega,t)  =  \frac{c_1\omega^s}{\ln \omega+\lambda_2-2} H_{1,2}^{2,0}\left[\frac{\omega^2}{4t^\alpha}\bigg|\begin{matrix}
    (1,\alpha) \\
    \left(-\frac{s}{2},1\right),  \left(-\frac{s}{2},1\right)
    \end{matrix}
    \right] \mbox{ for } t>0 \mbox{ and } 0<\alpha<2, 
\end{equation}
\begin{equation}
\label{uuu2}
    u(\omega,t)  = \frac{\omega^{s-2}t^\alpha}{\ln \omega+\lambda_2-2}\sum_{k=1}^n \frac{c_k\omega^{\frac{2k}{\alpha}}}{t^k}{}_3\Psi_1\left[\frac{4t^\alpha}{\omega^2}\bigg|\begin{matrix}
\left(1-\frac{k}{\alpha}-\frac{s}{2},1\right),\left(1-\frac{k}{\alpha}-\frac{s}{2},1\right),(1,1)\\
(1+\alpha-k, \alpha)
    \end{matrix}
    \right]  
\end{equation}
for $\alpha\ge 2$. Here $n \in \mathbb{N}$ satisfying $0\le n-1<\alpha\le n$ and $c_k$ ($k=1,\ldots,n$) are real constants. 

We note that the above solutions of the fractional diffusion-wave Eq.~(\ref{1}) can be expressed in terms of the exponential and Gauss hypergeometric functions when $\alpha=1$ (classical diffusion) and $\alpha=2$  (classical wave). 
By setting $\alpha=1$ and $s=-2$ in (\ref{uuu1}), we get the following solution through the formula (1.125) of \cite{Hnom} 
$$u(\omega,t)=\frac{c_1}{4\left(\ln{\omega}+\lambda_2-2\right)t}\exp{\left(-\frac{\omega^2}{4t}\right)}.$$
If we set $\alpha=2$ in (\ref{uuu2}), then it becomes
\begin{multline*}
    u(\omega,t)  = \frac{c_1\omega^{s-1}t}{\ln \omega+\lambda_2-2}{}_3\Psi_1\left[\frac{4t^2}{\omega^2}\bigg|\begin{matrix}
\left(\frac{1}{2}-\frac{s}{2},1\right),\left(\frac{1}{2}-\frac{s}{2},1\right),(1,1)\\
(2, 2)
    \end{matrix}
    \right]\\
    +\frac{c_2\omega^{s}}{\ln \omega+\lambda_2-2}{}_3\Psi_1\left[\frac{4t^2}{\omega^2}\bigg|\begin{matrix}
\left(-\frac{s}{2},1\right),\left(-\frac{s}{2},1\right),(1,1)\\
(1, 2)
    \end{matrix}
    \right].
\end{multline*}
By virtue of the following formulas
\begin{eqnarray*}
{}_3\Psi_1\left[z\biggr\vert\begin{array}{c}
(A_1,1), (A_2,1), (1,1)\\
(1,2)
\end{array}\right]&=&\Gamma(A_1)\Gamma(A_2){}_2F_1\left(\begin{array}{c}
A_1,A_2\\
\frac{1}{2}
\end{array}; \frac{z}{4}\right)\text{ for } |z|<4\mbox{ and}\\ 
{}_3\Psi_1\left[z\biggr\vert\begin{array}{c}
(A_1,1), (A_2,1), (1,1)\\
(2,2)
\end{array}\right]&=&\Gamma(A_1)\Gamma(A_2){}_2F_1\left(\begin{array}{c}
A_1,A_2\\
\frac{3}{2}
\end{array}; \frac{z}{4}\right) \text{ for }|z|<4,
\end{eqnarray*}
this solution equals to
\begin{multline*}
        u(\omega,t)  = \frac{c_1\Gamma\left(\frac{1}{2}-\frac{s}{2}\right)^2\omega^{s-1}t}{\ln \omega+\lambda_2-2}{}_2F_1\left(\begin{array}{c}
\frac{1}{2}-\frac{s}{2},\frac{1}{2}-\frac{s}{2}\\
\frac{3}{2}
\end{array}; \frac{t^2}{\omega^2}\right)\\
    +\frac{c_2\Gamma\left(-\frac{s}{2}\right)^2\omega^{s}}{\ln \omega+\lambda_2-2}{}_2F_1\left(\begin{array}{c}
-\frac{s}{2},-\frac{s}{2}\\
\frac{1}{2}
\end{array}; \frac{t^2}{\omega^2}\right).
\end{multline*}
For the following cases, invariant solutions of Eq.~(\ref{eqnew1}) are obtained analogously to case \ref{subsec:V11}.

\subsection{The generator $V_{2}=\left(s+\frac{1-\lambda_2}{2}\right)X_1+X_3$, $s\in \mathbb{R}$}
\textcolor{white}{.}\\ 
Invariant solutions of Eq.~(\ref{eqnew1}) with $\overline{c}(\omega)=\dfrac{\left(\lambda_2+1\right)\left(\lambda_3\omega^{\lambda_2}-\lambda_2+1\right)}{\omega\left(\lambda_3\omega^{\lambda_2}+\lambda_2+1\right)}$ are
\begin{equation}
\label{uuu3}
    u(\omega,t)  =  \frac{c_1\omega^s}{\lambda_3\omega^{\lambda_2}+\lambda_2+1} H_{1,2}^{2,0}\left[\frac{\omega^2}{4t^\alpha}\bigg|\begin{matrix}
    (1,\alpha) \\
    \left(-\frac{s}{2},1\right),  \left(-\frac{s}{2}+\frac{\lambda_2}{2},1\right)
    \end{matrix}
    \right] 
\end{equation}
for $t>0$ and $0<\alpha<2$, 
\begin{multline}
\label{uuu4}
    u(\omega,t)  = \frac{\omega^{s-2}t^\alpha}{\lambda_3\omega^{\lambda_2}+\lambda_2+1}\\
    \times\sum_{k=1}^n \frac{c_k\omega^{\frac{2k}{\alpha}}}{t^k}{}_3\Psi_1\left[\frac{4t^\alpha}{\omega^2}\bigg|\begin{matrix}
\left(1-\frac{k}{\alpha}-\frac{s}{2},1\right),\left(1-\frac{k}{\alpha}-\frac{s}{2}+\frac{\lambda_2}{2},1\right),(1,1)\\
(1+\alpha-k, \alpha)
    \end{matrix}
    \right] 
\end{multline}
for $\alpha\ge 2$.

Furthermore, if $\alpha=1$ then by setting  $s=\lambda_2-2$ and $s=-2$ in \eqref{uuu3}, we obtain the following solutions, respectively,
$$u(\omega,t)=\frac{4^{\frac{\lambda_2}{2}-1}c_1t^{\frac{\lambda_2}{2}-1}}{\lambda_3\omega^{\lambda_2}+\lambda_2+1}\exp{\left(-\frac{\omega^2}{4t}\right)}$$
and
$$u(\omega,t)=\frac{4^{-\frac{\lambda_2}{2}-1}c_1\omega^{\lambda_2}t^{-\frac{\lambda_2}{2}-1}}{\lambda_3\omega^{\lambda_2}+\lambda_2+1}\exp{\left(-\frac{\omega^2}{4t}\right)}.$$
By setting $\alpha=2$ in (\ref{uuu4}), we get the following solution
\begin{multline*}
    u(\omega,t)  = \frac{c_1\Gamma{\left(\frac{1}{2}-\frac{s}{2}\right)}\Gamma{\left(\frac{1}{2}-\frac{s}{2}+\frac{\lambda_2}{2}\right)}\omega^{s-1}t}{\lambda_3\omega^{\lambda_2}+\lambda_2+1}{}_2F_1\left(\begin{array}{c}
\frac{1}{2}-\frac{s}{2},\frac{1}{2}-\frac{s}{2}+\frac{\lambda_2}{2}\\
\frac{3}{2}
\end{array}; \frac{t^2}{\omega^2}\right)\\
    +\frac{c_2\Gamma\left(-\frac{s}{2}\right)\Gamma\left(-\frac{s}{2}+\frac{\lambda_2}{2}\right)\omega^{s}}{\lambda_3\omega^{\lambda_2}+\lambda_2+1}{}_2F_1\left(\begin{array}{c}
-\frac{s}{2},-\frac{s}{2}+\frac{\lambda_2}{2}\\
\frac{1}{2}
\end{array}; \frac{t^2}{\omega^2}\right).
\end{multline*}

\subsection{The generator $V_{3}=\left(s-\frac{\lambda_2}{2}\right)X_1+X_5$, $s\in \mathbb{R}$}
\textcolor{white}{.}\\ 
Invariant solutions of Eq.~(\ref{eqnew1}) with $\overline{c}(\omega)=\dfrac{\lambda_2\left(\epsilon e^{\lambda_2\omega}-1\right)}{\epsilon e^{\lambda_2\omega}+1}$ are
\begin{equation}
\label{uuu8}
    u(\omega,t)  = \frac{e^{s\omega}}{\epsilon e^{\lambda_2\omega}+1} \sum_{k=1}^n c_kt^{\alpha-k}E_{\alpha,1+\alpha-k}(s(s-\lambda_2)t^{\alpha}). 
\end{equation}

If we take $\alpha=1$ and $\alpha=2$ in (\ref{uuu8}), then we get the following solutions, respectively, 
$$u(\omega,t)=\frac{e^{s\omega}}{\epsilon e^{\lambda_2\omega}+1}e^{s(s-\lambda_2)t}\quad\text{and}$$
$$u(\omega,t)=\frac{e^{s\omega}}{\epsilon e^{\lambda_2\omega}+1}\left(\frac{c_1}{\sqrt{s(s-\lambda_2)}}\sinh{\left(\sqrt{s(s-\lambda_2)}t\right)}+c_2\cosh{\left(\sqrt{s(s-\lambda_2)}t\right)}\right).$$

\subsection{The generator $V_{4}=sX_1+X_6$, $s\in \mathbb{R}$}
\textcolor{white}{.}\\ 
Invariant solutions of Eq.~(\ref{eqnew1}) with $\overline{c}(\omega)=\lambda_2\tan\left(-\frac{\lambda_2}{2}\omega\right)$ are
\begin{equation}
\label{uuu5}
    u(\omega,t)  = \frac{e^{s\omega}}{\cos(\frac{\lambda_2}{2}\omega)} \sum_{k=1}^n c_kt^{\alpha-k}E_{\alpha,1+\alpha-k}\left(\left(\frac{\lambda_2^2}{4}+s^2\right)t^{\alpha}\right). 
\end{equation}

If we take $\alpha=1$ and $\alpha=2$ in (\ref{uuu5}), then we get the following solutions, respectively, 
$$u(\omega,t)=\frac{c_1e^{s\omega}}{\cos(\frac{\lambda_2}{2}\omega)}\exp{\left(\left(\frac{\lambda_2^2}{4}+s^2\right)t\right)}\quad\text{and}$$
$$u(\omega,t)=\frac{e^{s\omega}}{\cos(\frac{\lambda_2}{2}\omega)}\left(\frac{2c_1}{\sqrt{\lambda_2^2+4s^2}}\sinh{\left(\frac{\sqrt{\lambda_2^2+4s^2}}{2}t\right)}+c_2\cosh{\left(\frac{\sqrt{\lambda_2^2+4s^2}}{2}t\right)}\right).$$

\subsection{The generator $V_{5} = X_1+\epsilon X_8$, $\epsilon=\pm 1$}
\textcolor{white}{.}\\ 
Invariant solutions of Eq.~(\ref{eqnew1}) with $\overline{c}(\omega)=0$ are 
\begin{equation}
\label{uuu6}
u(\omega,t)=e^{\epsilon\omega}\sum_{k=1}^n c_kt^{\alpha-k}E_{\alpha,1+\alpha-k}(t^\alpha).
\end{equation}

By putting $\alpha=1$ and $\alpha=2$ in (\ref{uuu6}), we obtain the following solutions, respectively, 
$$u(\omega,t)=c_1e^{\epsilon\omega}e^{t}\quad\text{and}\quad u(\omega,t)=e^{\epsilon\omega}\left(c_1\sinh{t}+c_2\cosh{t}\right).$$

\subsection{The generator $V_{6}=X_1+\epsilon X_9$, $\epsilon=\pm 1$} 
\textcolor{white}{.}\\ 
Invariant solutions of Eq.~(\ref{eqnew1}) with $\overline{c}(\omega)=\dfrac{2 }{\omega}$ are 
\begin{equation}
\label{uuu7}
    u(\omega,t) = \frac{e^{\epsilon\omega}}{\omega}\sum_{k=1}^n c_kt^{\alpha-k} E_{\alpha, 1+\alpha-k}(t^\alpha).
\end{equation}
If we take $\alpha=1$ and $\alpha=2$ in (\ref{uuu7}), then we have the following solutions, respectively, 
$$u(\omega,t)=c_1\frac{e^{\epsilon\omega}}{\omega}e^{t}\quad\text{and}\quad u(\omega,t)=\frac{e^{\epsilon\omega}}{\omega}\left(c_1\sinh{t}+c_2\cosh{t}\right).$$

\subsection{The generator $V_{7}=X_9$}
\textcolor{white}{.}\\ 
Finally, invariant solutions of Eq.~(\ref{eqnew1}) with $\overline{c}(\omega)=\dfrac{2 }{\omega}$ are
\begin{equation*}
    u(\omega,t)=\frac{1}{\omega}\sum_{k=1}^n c_k t^{\alpha-k}.
\end{equation*}

\section*{Conclusions}
Through the application of Lie symmetry analysis, we have investigated a class of linear diffusion-wave equations with variable coefficients. As a result, we provide group classifications of the equations and derive exact invariant solutions that correspond to infinitesimal symmetries. These solutions can be considered as generalizations of the previously known solutions of the corresponding diffusion and wave equations (in means of order of time differentiation). 

Generally, finding solutions of reduced equations of fractional diffusion-wave equations are more complicated than finding solutions of reduced equations of diffusion ($\alpha=1$) and wave ($\alpha=2$) equations. For example, in G.~Bluman and S.~Kumei \cite{bluman1}, the reduced equations of wave equations have solutions that are expressed in terms of hypergeometric functions. 
In our case, we give solutions of the reduced equations in terms of generalized Wright functions and Fox H-functions, which can be considered as generalizations of hypergeometric functions.

\subsection*{Acknowledgements}
 
 This work was partially supported by the National University of Mongolia (Grant No.P2020-3966).

\bibliographystyle{amsplain}

\end{document}